# Forced dynamic dewetting of structured surfaces: Influence of surfactants

*Franziska Henrich[a], Dorota Linke[a], Hans Martin Sauer[b], Edgar Dörsam[b], Steffen Hardt[c], Hans-Jürgen Butt[a], Günter K. Auernhammer[a,d]*

[a] Max Planck Institute for Polymer Research, Ackermannweg 10, 55128 Mainz, Germany
[b] Technische Universität Darmstadt, Institut für Druckmaschinen und Druckverfahren, Magdalenenstr. 2, 64289 Darmstadt, Germany
[c] Technische Universität Darmstadt, Fachgebiet Nano- und Mikrofluidik, Alarich-Weiss-Straße 10, 64287 Darmstadt, Germany
[d] Leibniz Institut für Polymerforschung, Hohe Straße 6, 01069 Dresden, Germany, email: auernhammer@ipfdd.de

ABSTRACT:
We analyse the dewetting of printing plates for gravure printing with well-defined gravure cells. The printing plates were mounted on a rotating horizontal cylinder that is half immersed in an aqueous solution of the anionic surfactant sodium 1-decanesulfonate. The gravure plates and the presence of surfactants serve as one example of a real-world dewetting situation. When rotating the cylinder, a liquid meniscus was partially drawn out of the liquid forming a dynamic contact angle at the contact line. The dynamic contact angle is decreased on a structured surface as compared to a smooth one. This is due to contact line pinning at the borders of the gravure cells. Additionally, surfactants tend to decrease the dynamic receding contact angle. We consider the interplay between these two effects. We compare the height differences of the meniscus on the structured and unstructured area as function of dewetting speeds. The height difference increases with increasing dewetting speed. With increasing size of the gravure cells this height difference and the induced changes in the dynamic contact angle increased. By adding surfactant, the height difference and the changes in the contact angle for the same surface decreased. We further note that although the liquid dewets the printing plates some liquid is always left in the gravure cell. At high enough surfactant concentrations or high enough dewetting speed, the dynamic contact angles in the structured surface approach those in flat surfaces. We conclude that surfactant reduces the influence of surface structure on dynamic dewetting.

## 1. Introduction

Wetting and dewetting of solid surfaces play an important role in many natural and technological contexts, ranging from the movement of water drops on wings of butterflies and birds [1, 2] to printing [3, 4] and cleaning. The latter is important in industrial processes, e.g. the manufacturing of chips on silicon wafers [5-8]. For an optimal performance of any of these processes a fundamental understanding of the wetting properties is essential.

For simple liquids on flat surfaces some understanding of static and dynamic wetting has been achieved [3, 9-11]. The contact angle at the three-phase contact line (contact line in short), where liquid, substrate and gas (or a second liquid) meet, is a very convenient parameter to characterise the wetting behaviour. In the static case, two contact angles have to be distinguished: the static advancing contact angle $\theta_a$, and the static receding contact angle $\theta_r$. These contact angles are unique properties of the specific substrate and the liquid-gas or liquid-liquid combination. For moving contact lines, dynamic aspects have to be considered in addition to that. Several theories explain the dynamic contact angle as a function of wetting or dewetting speed of the contact line on various length scales. Two prominent examples are the molecular kinetic theory [12, 13] or the hydrodynamic theory [14-16]. A detailed description and comparison of the theories can be found in recent review articles [3, 9].

These modelling attempts went hand in hand with a wealth of experimental work on the wetting behaviour of single component liquids [3, 9, 17] Also, simulations on the wetting of single component liquids have been executed [10, 18, 19]. However, the dynamic wetting of multi-component liquids, like surfactant solutions, is less understood. In recent years, an increasing amount of research has been done on this topic [3, 20-27]. Surfactant molecules absorb at the interface and change the interfacial tension [28]. Flow-induced heterogeneities of the absorbed surfactant layer at the interface also influence the dynamic behaviour of the liquid. Concentration gradients at the interface lead to Marangoni stresses that change the hydrodynamic boundary condition [22, 23] or the flow profile inside the liquid [24, 29]. The majority of the studies investigate spontaneous wetting, e.g., the spontaneous spreading of drops. Less investigated is the forced wetting behaviour, e.g.., wetting and dewetting with a prescribed speed of the contact line. In the latter case, liquid moves over a solid surface due to external forces.

Previous work shows that surfactants influence the wetting behaviour even well below the critical micelle concentration (CMC) [24-27, 30, 31]. All of these studies have been performed on relatively smooth surfaces. However, in many technical wetting applications, the surfaces are not smooth but have a topography. Therefore, the question arises: How do surface structures influence the effect of surfactants on dynamic dewetting? Forced dewetting on structured surfaces has not been intensely studied so far [32-35].

In many technical processes (like printing processes), forced wetting on structured surfaces plays a major role. Transfer and dosing of printing inks and coating liquids is controlled by the filling and emptying of the engraved patterns of the printing forms and cylinders with the respective liquids. It is known that surfactants play a significant role here [36].

We aim for a better understanding of how surfactants influence forced wetting on structured surfaces. Here we combine our previous work on the dewetting of surfactant solutions on flat surfaces [24-27, 30, 31] with the effects of pinning on surface structures [32-35]. For a good comparison, our samples have structured and unstructured regions side-by-side. This allows for a direct comparison of how the surface structure and surfactants influence dewetting. For this purpose, we mounted different custom-made gravure printing plates on a specially designed rotating drum setup, where the drum is half immersed in the liquid. The gravure cells in the printing plate act as pinning sites for the receding contact line. This pinning strongly influences the dewetting behavior. We find that, with increasing surfactant concentration, the dewetting behavior of rough surfaces approaches that of smooth surfaces. From this we conclude that hydrodynamic and Marangoni effects close to the contact line dominate over pinning effects for high enough dewetting speeds.

## 2. Materials and Methods

### 2.1. Structured printing plates

We used custom-made gravure printing plates carrying a number of engraved cell raster areas with different cell geometries. The printing plates were provided by GT&W, Rödermark, Germany. The cells were mechanically engraved in a copper layer that has been galvanically deposited on 0.4 mm stainless steel sheets. After removing the grates, the engraved surface was coated by approximately 5 µm of electrochemically deposited polished chromium, which was then hardened. The microstructures were analysed with optical 3D surface microscopy (Nanofokus, 50 x magnification, Figure 1a). The measured dimensions of the gravures are summarized in Table 1. The inner side walls of all gravure cells on all printing plates are inclined by an identical angle of about 20° relative to the flat surface.

There is also a roughness on the unstructured areas between the structured areas due to the production process. As measured with atomic force microscopy, the roughness is approximately $R_g \approx 35\ nm$. The structured areas were 15 x 60 mm$^2$ in size and surrounded by an unstructured region (Figure 1b). Before using them for the wetting experiments, the plates were cleaned with Ethanol and flowing ultra-pure water.

Due to the production process, structured and unstructured areas thus chemically consist of the same material (electronically deposited chromium) and can be assumed to have the same wetting properties. In the experiments, we used the unstructured areas as an internal reference to analyze the influence of the structure on the dewetting behavior.

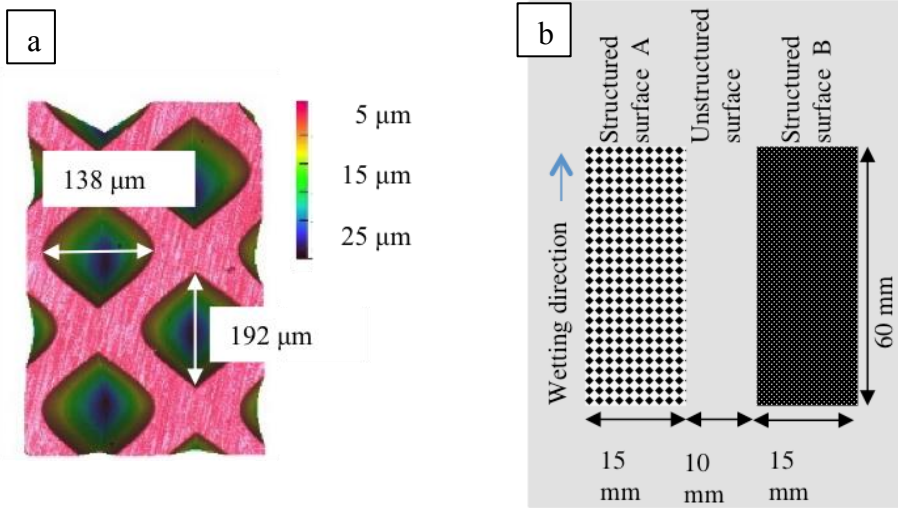

Figure 1: [a] Optical 3D surface microscopy image of the structured surface B-25. [b] Sketch of the printing plates with two structured areas surrounded by unstructured surface. The orientation of the structured areas on the printing plate is also visible in Figure 2b.

Table 1: Properties of the different areas on the printing plate. The labels reflect the distance between the neighbouring gravure cells and their depth.

| Label | Gravure cell dimensions | | | Distance between cells | |
|---|---|---|---|---|---|
| | $x$ [μ$m$] | $y$ [μ$m$] | Depth [μ$m$] | $x$ [μ$m$] | $y$ [μ$m$] |
| B-7 | 85 | 45 | 7 | 262 | 262 |
| B-9 | 98 | 62 | 9 | 262 | 262 |
| B-13 | 117 | 84 | 13 | 262 | 262 |
| B-25 | 192 | 138 | 25 | 262 | 262 |
| S-4 | 46 | 31 | 4 | 218 | 212 |
| S-7 | 63 | 49 | 7 | 218 | 212 |
| S-9 | 82 | 68 | 9 | 218 | 212 |
| S-19 | 135 | 121 | 19 | 218 | 212 |

### 2.2. Rotating drum setup

The rotating drum setup consists of a stainless-steel drum, which is placed in a closed stainless-steel container (Figure 2a). Seven windows allow for optical observation of the advancing and receding contact lines. The inside of the container has a height of 150 mm, a length of 170 mm and a depth of 90 mm. The windows are mounted at the front and backside of the container, one at the top and two at every side of the drum axis. The container is made watertight using Teflon tape in every joint. In this way, we avoid any contamination of the liquid due to the sealing and allow for a thorough cleaning of the setup. The axis of the drum is sealed using gland packing sealing rings, made of Teflon. All measurements are carried out in a water-saturated atmosphere.

After filling the container with the solution, we rotate the drum for at least 20 minutes to allow the air to be saturated with water vapour. To avoid surfactant transfer along the liquid surface from the receding to the advancing side, the container was filled to the axis of the drum [27], which corresponds to one liter of liquid.

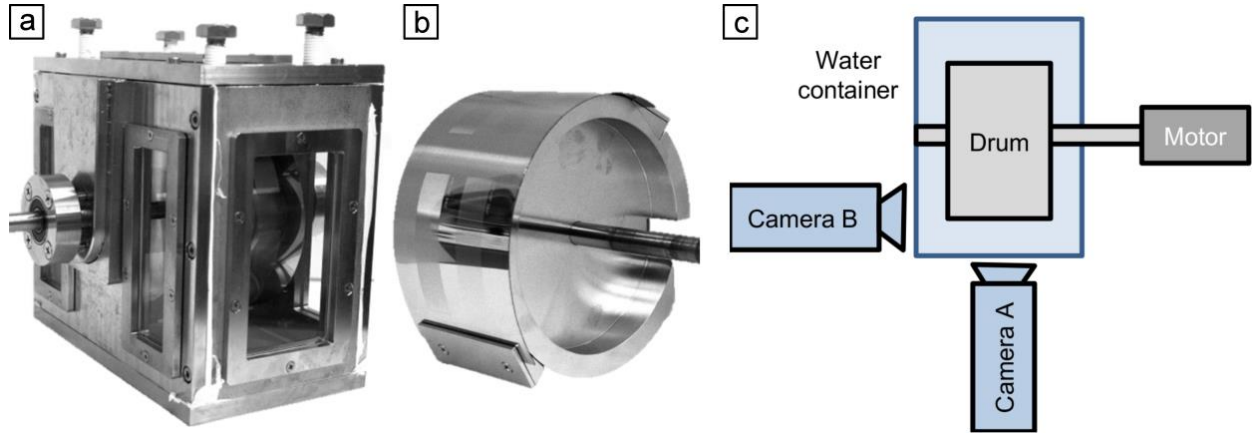

Figure 2: [a] Photograph of the rotating drum setup; [b] Photograph of the drum with a printing plate mounted on the outer curved surface of the drum. The structured areas on the printing plate are visible by their stronger scattering of light. [c] Sketch of the experimental setup, camera A is used for the height difference measurements, camera B for the contact angle measurements.

The cylindrical drum has a diameter of 120 mm, a width of 60 mm, and has the option of mounting different kinds of flat bendable or flexible plates on the cylinder (Figure 2b). In this work, the above described gravure printing plates were mounted on the drum. The drum is connected to a motor to vary the velocity of the printing plates between 0.1 mm/s and 100 mm/s. This motor allows a smooth motion of the drum. We use four different motors equipped with suitable gearings to cover one decade in velocity per motor. For comparison measurements on a smooth surface were performed, where a drum with a spherical segment geometry (coated with polystyrene) as described in [24] was used.

We observed the receding contact line with a high-speed camera (Photron, Fastcam SA-1), see Figure 2c. The camera was equipped either with a $2 \times$ magnification objective with a working distance of 35 mm and an illumination through the objective (for contact line observation, camera position A), or with a $12 \times$ magnification objective with a working distance of 300 mm and back light illumination (for contact angle observation, camera position B). Typical frame rates varied between 250 and 500 frames per second. Since the printing plates only allow bending in one direction, i.e., they take a cylindrical shape upon bending, side view imaging of the contact line is difficult. The rotating cylinder is blocking part of the view field of the side-view objective. This limits the optical resolution in side view. Additionally, the unstructured and the structured areas both along the optical path in side view. Consequently, the menisci of the structured and unstructured parts overlap on side-view images and contact angles cannot reliably measured for the structured areas using side-view imaging. It is only possible to measure the contact angle on the unstructured regions of the printing plates.

Instead we measured the height difference $\Delta h$ of contact line on the structured and unstructured areas as a function of dewetting speed (Figure 3). This height difference gives a direct measure of the changes in wetting properties due to the surface structures. As we will discuss later, this gives only indirect access to the overall height of the meniscus above the level of the unperturbed water surface.

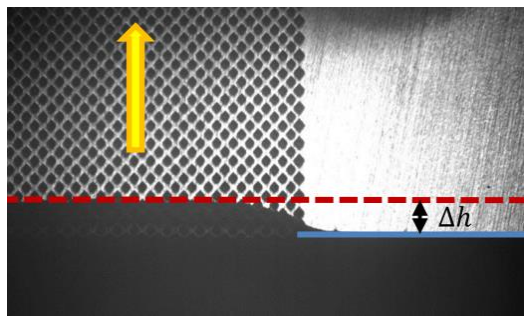

Figure 3: Image of the contact line. The red dashed line marks the contact line on the structured surface and the blue solid line on the unstructured surface. The difference between these is defined as the height difference $\Delta h$ (black arrow). The yellow arrow indicates the direction of motion of the surface.

### 2.3. Surfactant solution

We used solutions of the anionic surfactant sodium 1-decanesulfonate (S-1DeS) in ultra-pure water. S-1DeS was purchased from Sigma-Aldrich and used without further purification. The ultra-pure water was prepared by using an Arium® pro VF/UF& DI/UV (Sartorius) at a resistivity of 18.2 MΩcm. The surface tension of these solutions has been published elsewhere [24]. All used concentrations are below the critical micelle concentration of S-1DeS, i.e., the surfactant is molecularly dissolved in water in the form of single molecules, and no formation of aggregates is expected.

### 2.4. Measurement procedure

For cleaning before the experiment, the whole setup was rinsed overnight with flowing tap water then immediately rinsed with flowing ultra-pure water for one hour to avoid calcification in the setup. To ensure the absence of water-soluble impurities in the setup, we systematically used ultra-pure water as the first wetting liquid. Only when the measured values for contact angle and meniscus height did not change over an interval of half an hour of continuous measurements, the setup was considered to be clean.

We measured the height difference $\Delta h$ for all surfaces first for pure water and then for increasing surfactant concentration. This was done by successively adding the corresponding amount of surfactant. After each addition of surfactant, the solution was stirred for at least 20 min by rotating the drum at 100 mm/s. S-1DeS was added stepwise to achieve concentrations of up to 45 % of the critical micelle concentration (45 %CMC). The CMC of S-1DeS was measured with a Wilhelmy plate tensiometer (Dataphysics, DCAT 11EC) to be 38.5 mmol/l [24].

# 3. Results and Discussion

## 3.1. Pure water

### 3.1.1. Static contact angles

To measure the static contact angle on the printing plate, a drop of 10 µl water was placed (using an OCA 35 contact angle measuring device, DataPhysics). When inflating and deflating the drop, pinning of the contact line happened at the rims of the gravures of the printing areas. Due to this pinning, the quasi-static advancing and receding contact angle depends on the exact position of the contact line on the printing plate. For the receding contact angle, we observed variations of more than 15° while the contact line receded quasi-statically, see Figure 4 for the B-25 surface. Typically, the contact line is pinned at a corner of a gravure cell. Therefore, the contact angle decreases while being pinned at the edge of the gravure cell and increases after jumping to the next row of gravure cells. This pinning phenomenon at edges is well known and for example described in [17, 37].

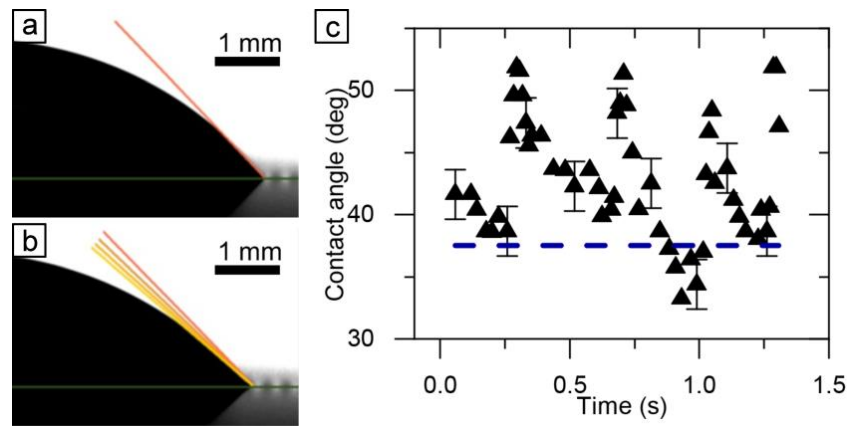

Figure 4: Receding contact angle oscillation of a deflating drop on the B-25 surface. During deflation of the drop the contact line moves to the left. [a] Contact angle on the flat part of the structured surface. [b] Due to pinning at the corner of the gravure cell, the contact angle changes while the drop moves over a gravure cell. The lines show the variation of the contact angle. [c] The contact angle variation during the movement of the contact line over the structured surface. Oscillations are due to pinning at the gravure cells. The contact angle varies by more than 15°. The dashed line is the contact angle on the smooth surface. The measurement error can be estimated to approximately 3°, as indicated in the plot.

Due to these pinning effects, it is important to take all measurements at a specific instant while the contact line moves over the printing plate. In the rest of the paper, we systematically took the measurements just before the contact line jumped from one row of gravure cells to the next row. As we already discussed above (section 2.2), side-view imaging to measure the contact angle on the structured part of the printing plate is not conclusive, because camera A either sees a superposition of different menisci on the structured and unstructured surfaces or only the unstructured part (but never the structured part alone). For this reason, we used the height difference $\Delta h$ between the structured and unstructured part as the primary parameter for the characterization of the dynamic wetting (Figure 3).

### 3.1.2. Dynamic measurements

For all different kinds of structures used in this work, the height difference $\Delta h$ increases with increasing velocity (Figure 5). While moving the plate out of the liquid container, the contact line is pinned at the top edges of the gravures until it unpins for all gravures in one line almost simultaneously. We systematically evaluated the height difference just before the contact line unpins. At a specific velocity (e.g., for the surface B-25 20 mm/s, red diamonds in Figure 5) the height difference reaches a plateau and further increases only for velocities above approximately 70 mm/s. The increase at higher velocities is associated with instabilities of the contact line. In that case the contact line is not straight anymore but starts to buckle. This instability is the first insinuation of the film formation that is observed at velocities above 80 mm/s. Therefore, no measurements are possible for speeds higher than 80 mm/s. For surfaces with shallow gravures (e.g. B-7) $\Delta h$ in this plateau regime is about the lateral distance between the gravure cells. For the surfaces with deeper gravures (e.g. B-25) the plateau value of $\Delta h$ amounts to more than twice this distance.

This implies that with increasing structure depth and width on the surface the influence of the structures on the dewetting of water increases. Since the inclination of the side walls is independent of the cell size (section 2.1), the variations in the height difference cannot be explained by a possible influence of the slopes of the side walls of the gravure cells. According to the standard pinning model [38] only the slope of the side walls but not the size of the gravure should influence pinning. We conclude that additional effects are involved, e.g., eventually slight but systematic changes in the in the local curvature of the rim of the gravures or by hydrodynamic effects induced by the emptying of the gravure cells (section 3.3.2).

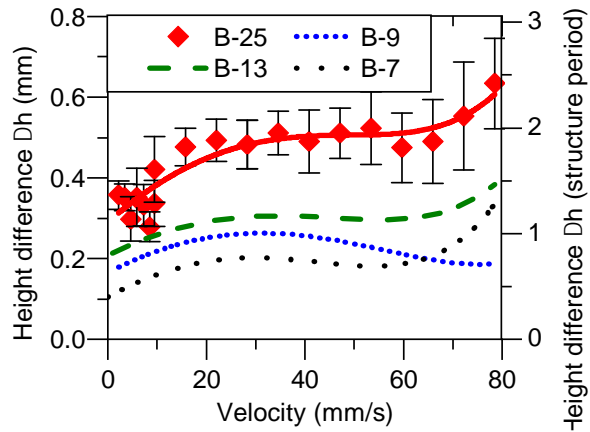

Figure 5: Measured height difference $\Delta h$ of the contact line for water between neighboring structured and unstructured parts of the printing plate (see Figure 3 for the definition of $\Delta h$). The experiments were performed on a printing plate with 262 µm x 262 µm distances between the gravure cells. The red diamonds represent the structured area with the deepest gravures. The dashed lines are guides to the eye for the data with smaller gravures (compare Table 1). The error bars represent the standard deviation of ten independent measurements.

As discussed above, direct optical determination of the dynamic receding contact angle on the structured areas is not easily possible. To overcome this limitation and to compare our results with previous experiments [23], we adopted the following strategy. We first measured the contact angle on the unstructured part of the surface. In regions with no structured areas along the line of sight this is possible in side view. The measurement results of the contact angle on the smooth part of the surface are shown in Figure 6a as violet open squared symbols. Due to suboptimal optical conditions, the uncertainty of the contact angle measurement is somewhat increased compared to the spherical segment drum [39]. The contact angle on the smooth surface decreases with increasing velocity.

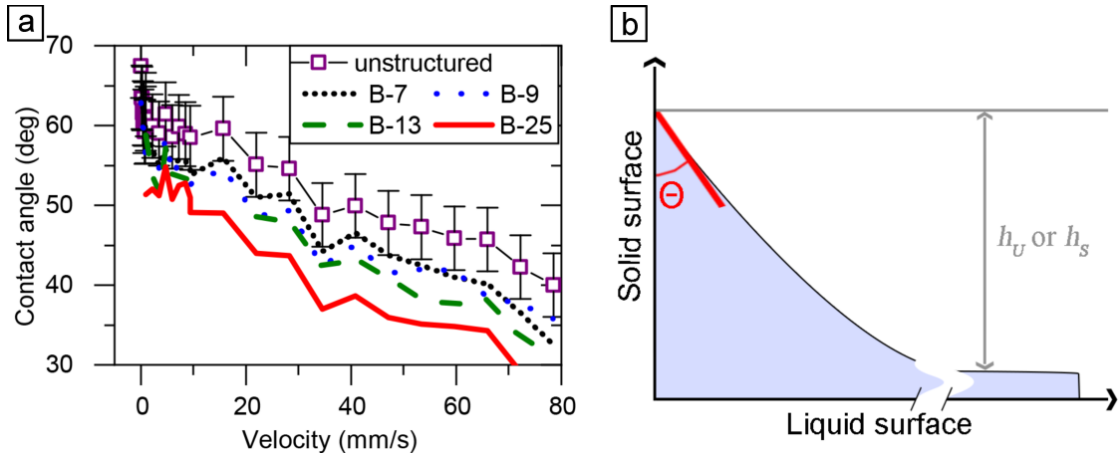

Figure 6: [a] Measured receding contact angles (opened squares) of water on different surfaces. To calculate the contact angle on the structured part of the plate, Eqs. (1) and (2) were used. Here, we used a surface tension of 72 mN/m and a density of 998.2 kg/m$^3$. For better visibility, the error bars are only shown for the unstructured surface. For the structured surfaces the error is ± 6°. [b] Height of the meniscus over the unperturbed liquid level. The *x*-axis is broken to sketch the whole meniscus shape.

To deduce the contact angle on the structured parts of the surface we used the fact that the shape of the meniscus follows the static shape for distances large compared to the slip length as well as the intermediate length scale [40]. Since we perform optical measurements with a resolution of around 100 µm, this is true for all our data. The height $h_u$ of the contact line over the unperturbed liquid level (Figure 6b) can be calculated by [41].

$$\gamma \sin \theta + \frac{1}{2} \varrho g h_u^2 = \gamma \tag{1}$$

In this expression, $\gamma$ is the surface tension of the liquid, $\theta$ the contact angle, $\varrho$ the density of the liquid, and $g$ is the gravitational acceleration. Using Eq. (1), we can calculate the height of the contact line over the unperturbed liquid level for the unstructured surface. By adding the height difference $\Delta h$ between the unstructured part and the different structured areas (Eq. (2)) we can calculate the height of the contact line over the unperturbed liquid level on the structured areas $h_s$.

$$h_s = h_u + \Delta h \tag{2}$$

Entering this value of $h_s$ in Eq. (1) instead of $h_u$, we obtain an estimate of the contact angle $\theta$ on the structured parts as well (Figure 6a, lines without symbols). In Figure 4 we have shown that the receding contact actually depended on the relative position of the contact to the gravures. When calculating the receding using we measured the minimal receding contact angle in the stick slip motion on the structured surfaces (Figure 4). The receding contact angle decreases with increasing velocity as well as with increasing structure depth on the surface (Figure 6).

### 3.2. Surfactant solutions

To figure out the influence of surfactants on the dewetting of structured surfaces we used concentrations of 15 %, 30 %, 45 % of the critical micelle concentration of S-1DeS. Like in the pure water case, we measured the height difference $\Delta h$ for increasing velocities until the contact line started to get unstable, i.e., for velocities up to the cross-over to film formation. As in the pure-water case, the height difference $\Delta h$ first increases with increasing velocity and then reached a plateau (Figure 7a). However, as shown in Figure 7b the height difference decreases with increasing surfactant concentration. Similar to [24, 26, 27] the film formation velocity as well as the contact angle decrease with increasing surfactant concentration (Figure 7c). The change in the contact angle increases with increasing surfactant concentration. This behavior is similar for all kinds of measured surfaces and comparable with the measurements of the same surfactant on a smooth surface [24] (Figure 7d).

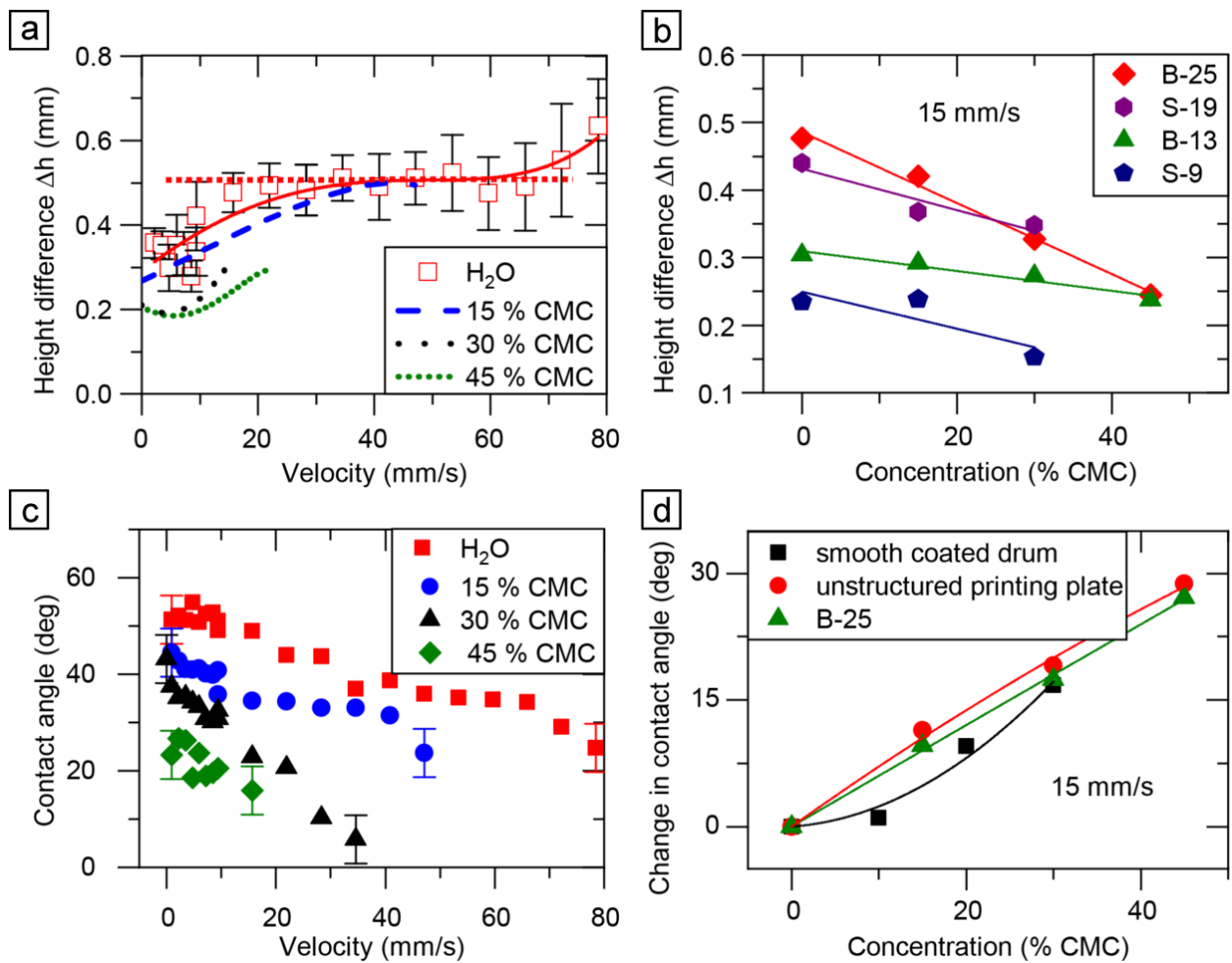

Figure 7: [a] Measured height difference of the contact line on the structured and unstructured part on the printing plate B-25 for different surfactant concentrations. The concentration varies from 0% to 45% of the critical micelle concentration of S-1DeS. The red dotted line illustrates the height difference plateau. The error bars are obtained from ten independent measurements. [b] Change in the height difference of the contact line on different structured plates for different concentrations at a velocity of 15 mm/s. The lines are guides to the eye. [c] Calculated contact angle based on Eqs. (1) and (2) on the surface B-25. The error bars are similar for all concentrations. [d] Comparison of the change in the contact angle due to addition of S-1DeS on the smooth part of the printing plate (red circles) and the structured surface B-25 (green rectangles) with the data published in [24] (black squares). The dashed lines are guides for the eye.

### 3.3. Hydrodynamic considerations

There are clear differences between the data measured on the smooth drum and the structured surfaces at low speeds and low surfactant concentrations (Figure 7d). For high surfactant concentrations and high speeds these differences tend to vanish. Our experimental results indicated that, the higher the contact line velocity and the higher the surfactant concentration, the more the dewetting seems to be dominated by hydrodynamic effects (e.g. Marangoni effects) and not by processes very close to the contact line, e.g., pinning of the contact line.

### 3.3.1. Surfactant-induced flow

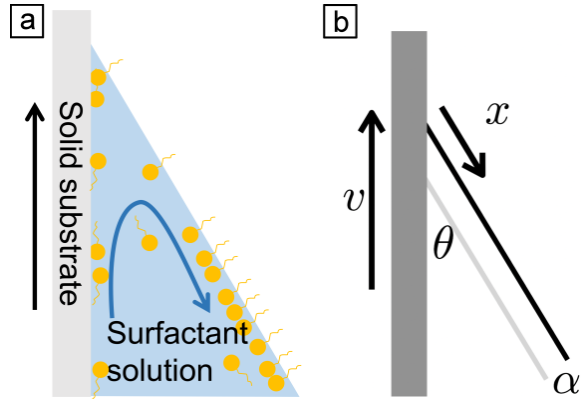

Figure 8: a) Sketch of the process close to the three-phase contact-line in the present of surfactants. The size of the surfactant molecules is not to scale (see main text for details). b) illustrates the definition of the quantities used in the scaling arguments.

For an explanation of this behaviour, we extend the model described earlier in [24, 26]. A sketch of the processes this model is based on is shown in Figure 8. Close to the substrate, the liquid moves with the solid surface towards the contact line. At the contact line the liquid changes its direction and flows along the liquid surface away from the contact line. This flow profile has been verified experimentally [24]. Correspondingly, close to the contact line fresh surface is continuously generated. Two mechanisms are possible to bring the surface concentration of the surfactant in equilibrium with the bulk of the liquid. Either surfactant molecules adsorbed to the substrate are transferred to the fresh liquid-air interface at the contact line, or surfactant molecules from the bulk of the liquid diffuse to the liquid-air interface. Presumably the adsorption of the surfactant at the solid surface is so strong (due to an electrostatic attraction between the surface and the surfactants) that no significant number of them can be transferred at the contact line to the liquid interface. For this reason, we assume that the dominating equilibration mechanism of the surface concentration of surfactant molecules is diffusion. As sketched in Figure 8, this leads to a region of lower surface concentration close to the contact line. The size of this region depends on two characteristic length scales that describe the equilibrium conditions and the advection-diffusion dynamics.

In equilibrium on can compare the surfactant concentration at the surface and in the bulk. A characteristic length scale $\alpha$ is the thickness of a liquid layer that contains as many surfactant molecules per unit area as the liquid-air interface. The thickness of this layer can be estimated using the Gibbs adsorption isotherm $\Gamma = -\frac{c}{2RT}\frac{d\gamma}{dc}$ and a linear dependence of the surface excess $\Gamma = \alpha c$ on the bulk concentration [26]. Here, $c$ is the bulk surfactant concentrations, $R$ the ideal gas constant and $T$ the absolute temperature. With an integration of the Gibbs adsorption isotherm between $c = 0$ and the CMC one obtains

$$\Delta\gamma = 2\alpha RT \cdot CMC, \tag{3}$$

where $\Delta\gamma$ is the surface tension difference between pure water and the surface tension at the CMC. For the S-1DeS, we measured a CMC of 38.5 mM and a surface tension of 39.7 mN/m at the CMC. This gives $\alpha \approx 180$ nm. This length scale is much smaller than the size of the gravure. But it is also much larger than molecular length scales, like the size of the surfactant molecules or length scales of the molecular motion at the contact line, like slip lengths or hopping distances. $\alpha$ is in fact a length scale at which hydrodynamic arguments can be assumed to hold. To refill the freshly generated surface with surfactant molecules, all the molecules from a region of $x = 2\alpha / \tan\theta$ are needed. The factor 2 is due to the assumed triangular shape of the region close to the contact line. At low velocities, the surface concentration of surfactant molecules is no longer in equilibrium in a region of this size close to the contact line. Note that the physical basis of the above argument is the conservation of the surfactant.

Diffusion brings the surfactant molecules to the liquid-air interface. This diffusion is approximately perpendicular to the streamlines lines of the hydrodynamic flow. The characteristic time scale can thus be estimated by the diffusive time scale

$$\tau_D = \frac{\alpha^2}{2D}, \tag{4}$$

with the diffusion coefficient of the surfactant $D = 6.98 \times 10^{-10}$ m$^2$/s. We obtain $\tau_D \approx 2.3$ µs. This time scale has to be compared to the advective time scale $\tau_A$ over $x = 2\alpha / \tan\theta$, $\tau_A = 2\alpha / (v \tan\theta)$, where $v$ is the flow velocity. For velocities at which $\tau_A$ is shorter than $\tau_D$ the region of non-equilibrium surface tension is expanded, because diffusion is not fast enough to replenish the liquid-air interface. This characteristic velocity is given by

$$v_A = 2D \frac{\tan\theta}{\alpha}. \tag{5}$$

For a typical contact angle of $\theta = 30°$, we obtain $v_A \approx 4.5$ mm/s. This implies that for almost all velocities used in our experiments the region of non-equilibrium surface tension is larger than $x = 2\alpha / \tan\theta$ and dominated by advection.

This concentration gradient along the liquid-air interface has important consequences on the flow profile. Due to the gradient in surface concentration there is also a gradient in surface tension along the liquid-air interface. This surface tension gradient and the associated Marangoni stress are counteracting the flow close to the contact line. This situation bears some similarities to the stagnant caps of bubbles rising in surfactant solutions [42, 43]. The flow-induced Marangoni stresses reduce the surfaces flow close to the contact line and confine the shear flow more or less to the bulk of the liquid. As we have observed in our previous work [24], this leads to an effectively increased viscous dissipation close to the contact line. The amount of this additional dissipation is a function of the surfactant concentration. In a simple picture, the surface tension varies between the value for pure water close to the contact line and the equilibrium value at the given surfactant concentration at a velocity-dependent distance from the contact line that measures in µm. The higher the surfactant concentration, the stronger this gradient.

This Marangoni-induced dissipative mechanism close to the contact line adds to the other mechanism that contribute to the dynamic contact angle. The experimental results of this and previous works [24, 26] show that, at high enough velocities, the effectively increased dissipation close to the contact line due to Marangoni stresses seems to dominate the motion of receding contact lines of surfactant solutions (see discussion of Figure 9 below). In our previous work, we observed a cross-over to a regime in which hydrodynamic models for dynamic receding contact angles are applicable.

### 3.3.2. Dependence on the characteristics of the surface structure

The effects of Marangoni stresses and pinning also show distinct differences, e.g. with respect to the velocity dependence and the dependence on the size of the gravure. The strength of the pinning at the gravure cells (as measured through the height difference $\Delta h$) depends on the size of the gravure (Figure 7b).

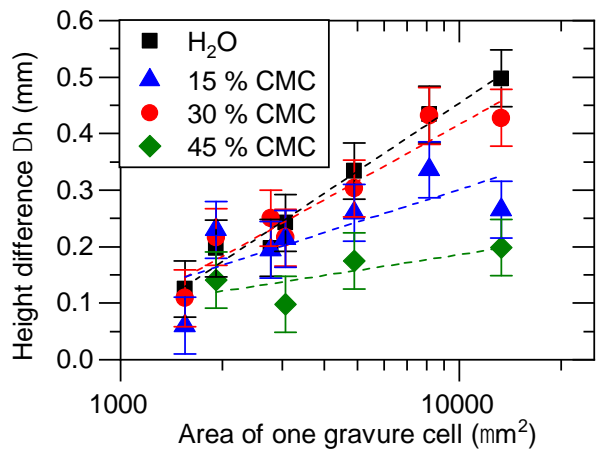

Figure 9: Measured height difference of the plateau (see Figure 7a) for all different kinds of surfaces employed. The height difference is plotted over the area of one gravure cell for both distances between the gravure cells. With increasing surfactant concentration, the influence of the structure decreases. The bars represent the standard deviation of the data points in the plateau. The dashed lines are guides to the eye.

A basic result of our measurements is the plateau in $\Delta h$ plotted against the dewetting velocity. We compare the plateaus in $\Delta h$ obtained for six different types of gravure cells (characterized by the gravure cell area) in Figure 9. The plateau values (see Figure 7a) is defined by calculating the average values of the data points between the initial rise of $\Delta h$ and its final increase at the speed of film formation.

For pure water, the height difference $\Delta h$ of the contact line increases with increasing size of the individual gravure cells (Figure 9, black squares). Actually, for all surfactant concentrations, the data points follow on average an increasing curve when plotting them against the size of the individual gravure cells. This illustrates that the size (e.g., the area) of the gravure cell is more important than the area fraction. Actually, the area fraction does not scale with the area of the cells, because we used different spacings between the gravure cells. We note that choosing the area of the gravures is an arbitrary choice, any other geometric feature of the single gravures that measures their size (length, depth, volume, etc.) would lead to the same conclusion.

When adding surfactant, the influence of the structure on the wetting behavior decreases with increasing concentration (compare section 3.3.2.). The higher the surfactant concentration, the smaller is the influence of the surface structure on the dewetting process (Figure 9). A similar effect has been observed in the contact angles that seem to converge to the same value independent of the surface structure for high enough surfactant concentrations (Figure 7).

### 3.3.3. Emptying of the gravure cells

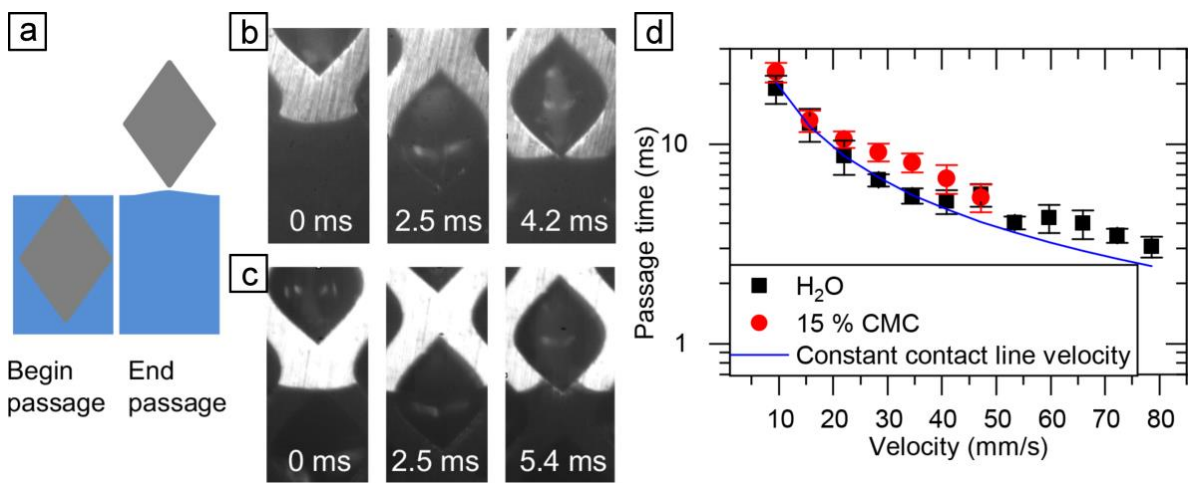

Figure 10: [a], [b], [c] show the emptying process of a gravure cell. While [a] defines the passage time as the time the contact line needs from the top corner of the gravure cell to depin from the bottom corner, [b] depicts the emptying for pure water and [c] for a solution with 15 %CMC of S-1DeS. The first frame in [b] shows the appearance of the gravure cell at the contact line, the second one the middle of the emptying process (after 2.5 ms), and the third frame the depinning of the contact line from bottom of the gravure cell. [d] presents the passage time for one gravure cell with and without surfactant for different velocities. The solid blue line indicates the passage time for a constant contact line velocity.

To rationalize why the height difference $\Delta h$ exhibits a plateau in a certain velocity range, we had a closer look at the emptying behaviour of single gravure cells. We imaged the process with a higher magnification objective (50x), which allows us to observe the emptying of single gravure cells (Figure 10). The passage time is defined as the time the contact line needs from the top corner of the gravure cell to depin from the bottom corner (Figure 10a). We measured the passage time for different velocities with and without surfactants in the solution. The passage time decreases with increasing velocity. For water, the passage time follows the prediction of a constant contact line velocity up to a velocity of 40 mm/s. Above this velocity, it increases above this value. This indicates that the contact line moves with different speeds over pinning sites, gravures, and the region between gravures (Figure 10d). This variation of the passage time correlates with the plateau in the height difference measurements (Figure 7). Similarly, for surfactant solutions the passage time decreases with increasing velocity. Above a velocity of 10 mm/s we observe deviations of the passage time from the values derived based on the given surface velocity. This indicates that the emptying behaviour of a single gravure cell differs for surfactant solutions and pure water. We attribute this difference to the additional dissipative mechanisms close to the contact line discussed above.

So far it is not clear whether the gravure cells will be completely empty after the passage of the contact line. Therefore, we had a closer look at the emptying mechanism. Due to optical reflection and the inclined side walls of the gravure cells, it is not possible to unambiguously decide if the gravure cells are completely empty after the contact line has depinned.

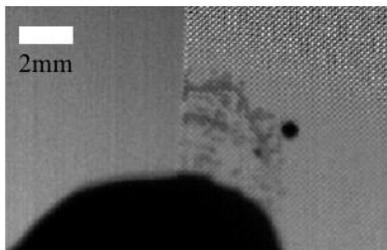

Figure 11: IR image of a moving drop. The left side shows an unstructured surface, the right side the structured surface B-25. Darker regions emit less IR radiation (e.g., are colder or reflect less IR radiation) than brighter regions.

To see if the emptying of the gravure cells is complete, we used an infrared camera (Infatec, VarioCam HD) to image the cells after emptying (Figure 11). We placed a 300 µl size water drop on a slightly tilted printing plate. The drop starts to move over the surface with an average velocity of approximately 10 mm/s. On the unstructured part, the temperature of the printing behind the drop is similar to the temperature in front of the drop. However, on the structured part, the temperature behind the drop is lower than in front of it. We attribute this temperature decrease to the evaporation of water that was left in the gravures after the passage of the contact line. The cells did not empty completely during the emptying/passage time. Note that the liquid left in the gravures pinches-off from the liquid in the bath. This process bears some similarities to the motion of contact lines over chemical heterogeneities as, e.g., discussed in [44].

To overcome the optical limitations and to see in detail what happens in the gravure cells we used a structured SU-8 substrate with circular holes (diameter 48 µm, depth 11 µm, distance between the centres 100 µm) and investigated the receding contact line on this surface. A drop is placed on the surface, and the volume is reduced by a syringe pump. Due to the decreasing volume the contact line moves over the surface and the receding contact line can be imaged. The average velocity of the contact line achieved that way was around 0.3 mm/s. Comparable to the movement of the receding contact line on the printing plates, the contact line pins at the edge of the holes and slips to the next hole afterwards. After the contact line has depinned from the cells there is still liquid inside the cells. More precisely, the evaporation of the water remaining in the cells typically takes 0.9 seconds.

Since all imaged cells of the SU8 surface show an incomplete emptying process, and additionally the IR-images of a moving drop over the structured printing plate show a difference in temperature, we conclude that some residual liquid remains inside the cells of the printing plate. This picture suggests that the measured passage time depends on the average contact line velocity as well as on the hydrodynamics of the emptying process. When the contact line depins from a (only partially emptied) gravure cell, the remaining liquid has to pinch off from the contact line. This leads to an additional hydrodynamic resistance related to the contact line motion. Especially, the pinch-off process of the contact line from the liquid remaining in the gravure cell can depend on the volume in the gravure cell. Thus, this additional hydrodynamic effect contributes to the pinning of the contact line and, especially, to the dependence of the pinning on the size of the gravure cells that was observed (Figs. 7 and 9).

## 4. Conclusion

On structured surfaces the dynamic contact angle decreases with increasing surfactant concentration. At a certain dewetting speed not a defined three-phase contact line but a continuous liquid film is formed. This speed, at which the contact line starts to get unstable and begins to buckle, decreases with increasing surfactant concentration. This behavior is qualitatively similar to the behavior on smooth surfaces [23, 24].

Usually structured surfaces show a strong pinning of the contact line. Here, we have shown in dynamic situations that the importance of pinning due to surface structures decreases with increasing surfactant concentration. With increasing surfactant concentration Marangoni stresses close to the contact line increase and start dominating the effect of pinning.

We demonstrated that the individual gravure cells are not completely empty after the contact line has passed over them. A small amount of liquid remains inside the cells. The remaining liquid has to pinch-off from the liquid in the receding contact line. This adds a dynamic mechanism of pinning that depends on the liquid volume in gravure cells, i.e., on their size.

## Acknowledgment

The authors thank Noemí Encinas García for the preparation of the SU8 surface and Ilka Hermes for the AFM measurements. We kindly acknowledge the financial support by the German Research Foundation (DFG) within the Collaborative Research Centre 1194 "Interaction of Transport and Wetting Processes", Project A02 (Auernhammer, Hardt), A06 (Henrich) and C03 (Butt).